# Optimization-in-the-Loop for Energy-Efficient 5G

F. Malandrino*, C. Casetti*, C.-F. Chiasserini*, G. Landi†

*: Politecnico di Torino, Torino, Italy – †: Nextworks s.r.l., Pisa, Italy

*Abstract*—We consider the problem of energy-efficient network management in 5G systems, where backhaul and fronthaul nodes have both networking and computational capabilities. We devise an optimization model accounting for the main features of 5G backhaul and fronthaul, and *jointly* solve the problems of (i) node switch on/off, (ii) VNF placement, and (iii) traffic routing. We implement an optimization module within an application on top of an SDN controller and NFV orchestrator, thus enabling swift, high-quality decisions based on current network conditions. Finally, we validate and test our scheme with real-world power consumption, network topology and traffic demand, assessing its performance as well as the relative importance of the main contributions to the total power consumption of the system.

## I. INTRODUCTION

The traditional distinction between computational facilities (e.g., servers) and networking equipment (e.g., switches) becomes blurred in 5G systems, where nodes can have both switching and computational capabilities. 5G backhaul and fronthaul nodes (hereinafter referred to as B/F nodes) can be seen as a pool of computational, storage and networking resources capable of running a set of virtual network functions (VNFs). The concatenation of VNFs, represented as a graph, defines services that are made available to higher layers or third parties.

In this context, applications are needed for monitoring the network traffic and energy consumption, as well as making and implementing complex decisions about: (i) which B/F nodes and links to activate, in order to reduce energy consumption; (ii) how many instances of each VNF are needed, and on which of the active B/F nodes they should run; (iii) how traffic should be routed between them. In traditional architectures, these problems could be studied separately: one could decide where to place servers (and which to activate in order to handle the incoming traffic load), under the assumption that the network infrastructure is given and immutable. Similarly, traditional network optimization typically takes as an input a traffic matrix, summarizing how much data has to be transferred between any two network nodes. These approaches are no longer effective, and sometimes not even viable, in 5G networks, where such decisions as activating or deactivating a B/F node impact both the traffic *processing* and *routing* capabilities of the network.

The situation is further complicated by the nature of traffic 5G networks are expected to serve. As exemplified in Fig. 1, traffic flows have to traverse a *logical* graph made up of multiple processing steps, each corresponding to a VNF. Such graph can be fairly complex and connected in a mesh-like fashion, and our task is to match it with the *physical* graph made of B/F nodes and links. In doing so, we need to account for the fact that traffic flows can trigger additional flows as they traverse VNFs. A typical example of a logical graph is the one implementing the virtual Evolved Packet Core (vEPC). Fig. 1 depicts user traffic captured by the Remote Radio Head (RRH), traversing the eNodeB (eNB) and going to the Packet/Service Gateway (P/S-GW). On its way, it generates additional traffic from the eNB to the Home Subscriber Server (HSS), through the Mobility Management Entity (MME). The result is that traditional flow conservation laws do not hold in 5G networks, as flows of different types get *transformed* into one another and then served individually.

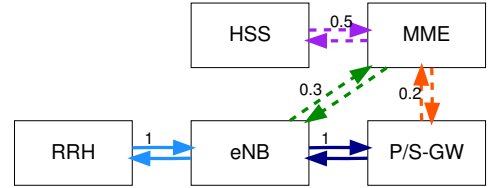

Fig. 1. Logical graph for vEPC. Solid lines correspond to user traffic, dashed lines to control traffic.

On the positive side, the software-defined nature of 5G networks affords us a significant opportunity to make their management more effective; indeed, to literally *optimize* them. Optimization is traditionally regarded as a useful technique to design networks, but of little use for their real-time management. In this paper, we depart from this vision and leverage the software-defined nature of 5G networks to make optimizers *interact* with SDN controllers and NFV orchestrators (NFVOs). These entities are indeed in a perfect position to swiftly implement real-time, high-quality decisions based on the outcome of optimization problems.

To achieve this goal, our paper provides the following main contributions:
- a *model* accounting for the complexity of both 5G networks and the traffic they have to serve;
- a *problem formulation* allowing us to make joint decisions on (i) B/F node activation, (ii) number and placement of the VNF instances we need, and (iii) traffic routing;
- a *solution strategy*, called OptiLoop, that uses optimization *in the loop* thus providing near-optimal decisions in *real-time*;
- an *implementation* of our solution in an OpenDaylight/OpenStack testbed as well as in an emulated testbed.

The remainder of this paper is organized as follows. We review related work in Sec. II. Next, we present our system model and problem formulation in Sec. III, and detail the OptiLoop solution strategy in Sec. IV. We then describe our testbeds' architecture, reference scenario and benchmarks in

Sec. V, present numerical results in Sec. VI, and conclude the paper in Sec. VII.

## II. RELATED WORK

Many works on VNF placement and traffic routing, including [1]–[3], take the approach of *matching* VNF and physical topology graphs, also proposing efficient solution strategies for the ensuing MILP problems. The optimization objectives are: minimizing network usage in [1], minimizing VNF deployment cost in [2], minimizing CAPEX and OPEX in [3]. The later work [4] takes an iterative approach, making VNF placement and routing decisions when a request arrives. [5] takes the VNF placement as given and focuses on scheduling and routing.

Other works focus on the interaction between mobile operators and third parties using their services. As an example, [6] considers a market where operators bid to serve incoming demands. Among energy-aware works, [7] seeks to optimize VNF placement and job scheduling in order to minimize energy consumption. However, the algorithm presented in [7] optimizes the server utilization but neglects the energy consumed by network elements such as B/F nodes.

Among the services that can be provided through SDN/NFV-based networks, a prominent example is the EPC. As suggested by the survey in [8], ILP and MILP are the most popular modeling tools, and heuristic algorithms the most popular solution strategy. A common theme [9]–[11] is splitting EPC elements, e.g., the Packet Gateway (P-GW) and Service Gateway (S-GW), into separate sub-elements, one dealing with control traffic and the other with user traffic. [12] finds that such an approach reduces the total cost of ownership. Interestingly, other works, e.g., [13], [14], take the opposite approach and merge P-GW and S-GW in a single entity (the P/S-GW). [13] focuses on the MME and proposes to implement it through four separate VNFs, whose number can vary so as to accommodate traffic fluctuations. Closer to our own effort is the recent work in [15], which studies the problem of placing the VNFs implementing the main EPC network functions – S-GW, P-GW and MME – across the available physical machines, subject to limits on their power and link capacity.

### A. Novelty

Our approach is novel with respect to the above works in several important ways:
1) first and foremost, the scope of our work: we jointly account for (i) the number and placement of VNF instances, (ii) traffic routing, and (iii) network management, e.g., activating/deactivating B/F nodes and links;
2) at the modeling level: accounting for the complexity of 5G traffic, with requests that originate at a network endpoint and traverse multiple VNFs, triggering additional requests as they do so (hence the quantity of traffic changes across processing steps);
3) as far as objectives are concerned: adopting energy-saving as our priority and using detailed and realistic energy models, instead of proxy metrics as in [7];
4) from a solution strategy viewpoint: optimizing in the loop, i.e., using optimization as a tool rather than a mere analysis technique;
5) at implementation level: validating and testing our approach through a testbed based on OpenDaylight and OpenStack.

## III. SYSTEM MODEL

Our model is based on two graphs, a logical one and a physical one. For simplicity, we describe it with reference to unidirectional traffic; notice however that our model and our results also account for bidirectional traffic. Tab. I summarizes all the notation we introduce below.

### A. The logical graph

The *logical* graph, exemplified in Fig. 1, describes where, i.e., which endpoint, the traffic comes from, and how it is processed. Its vertices are either *endpoints* $e \in \mathcal{E}$ or *VNFs* $v \in \mathcal{V}$. With reference to Fig. 1, we have $\mathcal{E} = \{\text{RRH}\}$, and $\mathcal{V} = \{\text{eNB}, \text{P/S-GW}, \text{MME}, \text{HSS}\}$.

On the logical graph, we have logical flows $l(e, v_1, v_2)$ representing data originating from endpoint $e$ and going from VNF $v_1$ to VNF $v_2$. Additionally, with an abuse of notation, we indicate with $l(e, v)$ flows that start from endpoint $e$ and are first processed at VNF $v$, e.g., from the RRH to the eNB in Fig. 1. Note that keeping track of the endpoint at which flows originate, i.e., having an $e$ index in our variables, serves a manifold purpose. First, it allows our model to account for the fact that different types of traffic (i.e., originating from different endpoints) may need different processing, i.e., traverse different VNF graphs. Furthermore, such VNF graphs may overlap; in this case, keeping track of the origin of the flows makes it possible to distinguish them even if they traverse the same VNF. Finally, it allows routing each flow in a different way, in both the logical and the physical graph. Notice that different traffic flows coming from the same physical endpoint can be distinguished by associating them to different *logical* endpoints.

Another important aspect of the system is that *there is no flow conservation in the logical graph*. As an example, in Fig. 1 we see a user flow of 1 traffic unit going from the RRH to eNB and thence to the gateway, which triggers some additional control traffic from the eNB and the gateway to the MME. Indeed, the following *generalized flow conservation* law holds for each endpoint $e$ and VNFs $v_2, v_3$:

$$l(e, v_2, v_3) = \sum_{v_1 \in \mathcal{V}} l(e, v_1, v_2)\chi(v_1, v_2, v_3) + l(e, v_2)\chi(e, v_2, v_3).$$

The above expression represents the logical flow originated at endpoint $e$, outgoing from VNF $v_2$ and directed to VNF $v_3$. Such a quantity is equal to the sum between logical flows entering $v_2$, from either a VNF $v_1$ or the endpoint $e$ itself, multiplied by a factor $\chi$. In particular, $\chi(v_1, v_2, v_3)$ is used to quantify the amount of logical flow directed to $v_3$ that is generated when traffic coming from $v_1$ is processed at VNF $v_2$. With reference to the eNB in Fig. 1, we have $\chi(\text{RRH}, \text{eNB}, \text{P/S-GW}) = 1$, while $\chi(\text{RRH}, \text{eNB}, \text{MME}) = 0.3$. Similarly, for the gateway, we have $\chi(\text{eNB}, \text{P/S-GW}, \text{MME}) = 0.2$. At the MME we have flow conservation, i.e., $\chi(\text{eNB}, \text{MME}, \text{HSS}) = \chi(\text{P/S-GW}, \text{MME}, \text{HSS}) = 1$. In $\chi(e, v_2, v_3)$, we abuse the notation and allow the first index of $\chi$ to be an endpoint instead of a VNF. We remark that $\chi$-values lower than one can also represent, e.g., a firewall dropping some of the incoming traffic. Also notice that $\chi$-values different from one can happen for both control traffic (e.g., the eNB in Fig. 1) and user traffic (as in the case of the firewall).

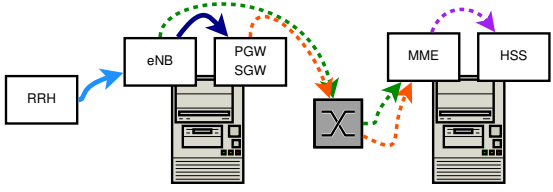

Fig. 2. Example implementation of the logical graph in Fig. 1 over a physical network. Each line corresponds to a physical flow, i.e., to a $\tau$-variable; their color and style match the logical flows in Fig. 1.

### B. The physical graph

In the physical graph, vertices correspond to the endpoints $e \in \mathcal{E}$ and the B/F nodes $c \in \mathcal{C}$. In general, B/F nodes have computational capabilities $k(c)$; B/F nodes that cannot host any VNF (e.g., switches) have $k(c) = 0$. Fig. 2 presents a possible implementation of the logical VNF graph in Fig. 1, where VNFs are placed on each of the two B/F nodes with processing capabilities. For simplicity, we present our model with reference to the case where multiple VNF instances can be deployed across different nodes, but at most one instance of each VNF can be deployed at each B/F node.

Physical links $(i, j) \in \mathcal{L} \subseteq (\mathcal{C} \cup \mathcal{E})^2$ have a capacity $C_{i,j}$, corresponding to the maximum amount of traffic that can go from B/F node $i$ to B/F node $j$. Traffic traversing link $(i, j)$ is also subject to a network delay $D_{i,j}$.

Our main variable is represented by *physical flows* $\tau_{i,j}(e, v_1, v_2)$, representing the amount of traffic that was originated from endpoint $e$, last visited VNF $v_1$, will next visit VNF $v_2$, and is now traveling on link $(i, j)$. Recall that we have to keep track of the flow originating endpint, in order to model traffic routing. If the flow has never been processed, i.e., it is going from $e \in \mathcal{E}$ to its first VNF $v \in \mathcal{V}$, we will conventionally set $v_1 = v_2 = v$ and write $\tau_{i,j}(e, v, v)$.

Given a B/F node $c \in \mathcal{C}$, we denote by $t_c(e, v_1, v_2)$ the amount of traffic that is just *transiting* by $c$ (i.e., it is *not* processed at $c$) and it was originated at $e$, last visited VNF $v_1$ and will next visit VNF $v_2$. Similarly, $p_c(e, v_1, v_2)$ is the traffic that *is processed* at B/F node $c$, it was originated at $e$, and last visited VNF $v_1$. Note that $p_c(e, v_1, v_2) > 0$ implies that an instance of VNF $v_2$ is deployed at $c$.

Traffic being processed at VNF $v$ is subject to a delay $D(v)$. Normally, processing delay is linked to the amount of resources (e.g., CPU) allocated to each VNF, and such an amount depends on the other VNFs deployed at the same B/F node. In our case, however, energy is the main metric of interest, and we can therefore assume that no VNF will be allocated more resources than the minimum amount required by the VNF itself.

A first constraint we need to impose is that, given a generic VNF $v_2$, the traffic originated at $e$, that has been processed through VNF $v_1$ and is entering B/F node $c$, is either (i) processed at an instance of $v_2$ located in $c$, or (ii) transiting by $c$ while being routed toward an instance of $v_2$. Thus, for any $c, e, v_1, v_2$, we have:

$$\sum_{(i,c)\in\mathcal{L}} \tau_{i,c}(e, v_1, v_2) = t_c(e, v_1, v_2) + p_c(e, v_1, v_2). \quad (1)$$

A similar constraint concerns the traffic outgoing from $c$. For any $c$, endpoint $e$ and VNFs $v_2, v_3$, we have:

$$\sum_{(c,j)\in\mathcal{L}} \tau_{c,j}(e, v_2, v_3) = t_c(e, v_2, v_3) + \sum_{v_1 \in \mathcal{V}} p_c(e, v_1, v_2)\chi(v_1, v_2, v_3) \quad (2)$$

where $v_2$ is the last VNF that traffic visited, either before arriving at $c$ (if traffic just transits by $c$) or at $c$ itself (if $v_2$ is deployed therein, i.e., $p_c(e, v_1, v_2) > 0$). $v_3$ instead is the VNF that traffic will visit next. In other words, (1)–(2) enforce *ordinary* flow conservation for the traffic that is transiting at $c$, i.e., using $c$ as a traditional switch, and *generalized* flow conservation for the traffic that is processed at $c$.

Next, we need to ensure that we only use active B/F nodes and links, and their capacity is not exceeded. We define two sets of binary variables, $x_{i,j}$ and $y_c$, indicating whether link $(i, j)$ and B/F node $c$ are active or not.

For links, we need to impose:

$$x_{i,j} \leq \min\{y_i, y_j\}, \quad \forall (i,j) \in \mathcal{L}, \quad (3)$$

i.e., no link can be active if either of its ends is off, and

$$\sum_{e \in \mathcal{E}} \sum_{v_1, v_2 \in \mathcal{V}} \tau_{i,j}(e, v_1, v_2) \leq x_{i,j} C_{i,j}, \quad \forall (i,j) \in \mathcal{L}. \quad (4)$$

With regard to processing, inactive B/F nodes cannot host any VNF. We track this through a binary variable $\delta(c, v)$ expressing whether an instance of VNF $v$ is deployed at B/F node $c$, and impose:

$$\delta(c, v) \leq y_c, \quad \forall c \in \mathcal{C}, v \in \mathcal{V}. \quad (5)$$

Additionally, no processing can be done for VNFs that are not deployed at a given B/F node:

$$p_c(e, v_1, v_2) \leq \delta(c, v_2) k(c), \quad \forall c \in \mathcal{C}, e \in \mathcal{E}, v_1, v_2 \in \mathcal{V}. \quad (6)$$

Finally, *each traffic unit* processed by VNF $v$ requires $r(v)$ computational capability, and, assuming $c$ is a software switch, each unit of traffic switched by $c$ consumes $\rho(c)$ CPU. Clearly, the computational capability of each B/F node $c$ must be sufficient for both, i.e., for any B/F node $c$,

$$\sum_{e \in \mathcal{E}} \sum_{v_1 \in \mathcal{V}} \sum_{v_2 \in \mathcal{V}} \left[ r(v_2) p_c(e, v_1, v_2) + \right.$$
$$\left. + \rho(c) \sum_{(c,j) \in \mathcal{L}} \tau_{c,j}(e, v_1, v_2) \right] \leq k(c), \quad (7)$$

where $\rho(c)$ multiplies the total traffic outgoing from $c$.

TABLE I
NOTATION

| Symbol | Type | Meaning |
|---|---|---|
| $\mathcal{E}$ | Set | Set of network endpoints |
| $\mathcal{C}$ | Set | Set of B/F nodes |
| $\mathcal{L}$ | Set | Set of links |
| $\mathcal{V}$ | Set | Set of VNFs |
| $C_{i,j}$ | Parameter | Capacity of link $(i,j) \in \mathcal{L}$ |
| $\chi(v_1, v_2, v_3)$ | Parameter | How much traffic resulting from the processing at VNF $v_2$, which was previously processed at VNF $v_1$, is meant to be next processed at VNF $v_3$ |
| $\delta(c,v)$ | Binary var. | Whether we deploy VNF $v \in \mathcal{V}$ at B/F node $c \in \mathcal{C}$ |
| $f_0$ | Function | Energy consumption due to placing a VNF at a B/F node |
| $f_{\text{idle}}$ | Function | Energy consumption due to activating a B/F node |
| $f_{\text{proc}}$ | Function | Traffic-dependent energy consumption due to processing |
| $f_{\text{sw}}, f_{\text{link}}$ | Function | Traffic-dependent energy consumption at switches and links |
| $k(c)$ | Parameter | Computational capability of B/F node $c \in \mathcal{C}$ |
| $l(e, v_1, v_2)$ | Parameter | Logical flow originated at $e \in \mathcal{E}$ and going from VNF $v_1 \in \mathcal{V}$ to VNF $v_2 \in \mathcal{V}$ |
| $l(e, v)$ | Parameter | Logical flow originating at $e \in \mathcal{E}$ and first being processed at VNF $v \in \mathcal{V}$ |
| $p_c(e, v_1, v_2)$ | Continuous variable | How much traffic coming from users connected to endpoint $e \in \mathcal{E}$ for service that was last processed at VNF $v_1$ is processed by an instance of VNF $v_2$ deployed at B/F node $c$ |
| $r(v)$ | Parameter | Computational capability required to process one traffic unit of VNF $v \in \mathcal{V}$ |
| $\rho(c)$ | Parameter | Computational capability consumed by one unit of traffic transiting by B/F node (SW switch) $c \in \mathcal{C}$ |
| $\tau_{i,j}(e, v_1, v_2)$ | Continuous variable | How much traffic coming from users connected to endpoint $e \in \mathcal{E}$ that was last processed at VNF $v_1$ and meant to be next processed at VNF $v_2$ goes through link $(i,j) \in \mathcal{L}$ |
| $t_c(e, v_1, v_2)$ | Continuous variable | How much traffic originating from $e$ that was last processed at VNF $v_1$ and meant to be next processed at VNF $v_2$ transits (without processing) by B/F node $c \in \mathcal{C}$ |
| $x_{i,j}$ | Binary var. | Whether link $(i,j) \in \mathcal{L}$ is active |
| $y_c$ | Binary var. | Whether B/F node $c \in \mathcal{C}$ is active |

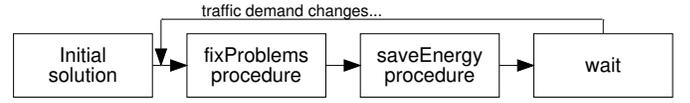

Fig. 3. The OptiLoop strategy. We begin by obtaining an initial feasible solution, as described in Sec. IV-A. After that, we periodically check the current solution for problems (procedure `fixProblems`, described in Alg. 1) and for opportunities to deactivate some B/F nodes and/or links (procedure `saveEnergy`, described in Alg. 2).

Next, we ensure that the delay of the traffic originated at any endpoint $e$ does not exceed a threshold $D^{\max}(e)$:

$$\frac{\sum_{i,j \in \mathcal{L}} \sum_{v_1, v_2 \in \mathcal{V}} D_{i,j} \tau_{i,j}(e, v_1, v_2)}{\sum_{v \in \mathcal{V}} l(e, v)} +$$
$$+ \frac{\sum_{v_1, v_2 \in \mathcal{V}} \sum_{c \in \mathcal{C}} D(v_2) p_c(e, v_1, v_2)}{\sum_{v \in \mathcal{V}} l(e, v)} \leq D^{\max}(e). \quad (8)$$

The two terms on the left hand side of (8) correspond to the network and processing delay, respectively.

At last, logical and physical flows have to match. To this end, it is sufficient to impose that, for each logical flow $l(e, v)$ going from endpoint $e$ to VNF $v$, there are corresponding physical flows of the type $\tau_{e,j}(e, v, v)$, such that:

$$l(e, v) = \sum_{(e,j) \in \mathcal{L}} \tau_{e,j}(e, v, v), \quad \forall e \in \mathcal{E}, v \in \mathcal{V}. \quad (9)$$

Eq. (9) ensures that the traffic injected from endpoints to B/F nodes on the physical graph matches the logical traffic going from endpoints to VNFs. Thanks to the flow conservation constraints (1)–(2), this also implies that such traffic is processed and transformed as dictated by the $\chi$-parameters, i.e., that all physical flows match their logical counterpart.

*C. Energy and objective*

There are five contributions to the overall energy consumption we are interested in tracking:

- activating a B/F node, resulting in a consumption of $f_{\text{idle}}$;
- placing a VNF on a B/F node, resulting in a consumption $f_0$ due to, e.g., virtual machines (VMs) or containers overhead;
- using said VNF, resulting in a consumption of $f_{\text{proc}}$ depending on the computational resources used;
- switching traffic at a B/F node, resulting in a consumption of $f_{\text{sw}}$ depending on the traffic switched by the node;
- having traffic going through links, resulting in a consumption of $f_{\text{link}}$ depending on the traffic over each link.

The total energy consumption due to the above four components is as follows:

$$E_{\text{idle}} = \sum_{c \in \mathcal{C}} f_{\text{idle}}(y_c); \qquad E_0 = \sum_{c \in \mathcal{C}} \sum_{v_2 \in \mathcal{V}} f_0(\delta(c, v_2));$$

$$E_{\text{proc}} = \sum_{c \in \mathcal{C}} f_{\text{proc}} \left( \sum_{v_2 \in \mathcal{V}} r(v) \sum_{e \in \mathcal{E}} \sum_{v_1 \in \mathcal{V}} p_c(e, v_1, v_2) \right);$$

$$E_{\text{sw}} = \sum_{c \in \mathcal{C}} f_{\text{sw}} \left( \sum_{e \in \mathcal{E}} \sum_{v_1, v_2 \in \mathcal{V}} \tau_{c,j}(e, v_1, v_2) \right);$$

$$E_{\text{link}} = \sum_{(i,j) \in \mathcal{L}} f_{\text{link}} \left( \sum_{e \in \mathcal{E}} \sum_{v_1, v_2 \in \mathcal{V}} \tau_{i,j}(e, v_1, v_2) \right).$$

Given all this, our objective can be written as:

$$\min_{x,y} E = E_0 + E_{\text{proc}} + E_{\text{idle}} + E_{\text{sw}} + E_{\text{link}}. \quad (10)$$

IV. THE OPTILOOP STRATEGY

The problem stated in Sec. III falls into the MILP category, and is thus impractical to solve in real time. We can however solve its *relaxed* version, where binary variables are allowed to take any value in $[0, 1]$. Optimal solutions to the relaxed models cannot be directly used to manage (or plan) a network; however, they can provide useful guidelines.

Our basic idea of is to leverage the software-defined nature of our network to make an optimizer *interact* with SDN controllers and NFVOs, i.e., optimize problems as a part of our network management strategy. Our solution strategy is called OptiLoop (for Optimization in the Loop) and it includes the following steps, as outlined in Fig. 3:

1) we initialize the system with a feasible (albeit potentially suboptimal) solution, as detailed in Sec. IV-A;
2) after that, we periodically:
   a) check that the network configuration is adequate to the current (and/or predicted) demand;
   b) if not so, activate additional VNFs, B/F nodes, and/or links as needed;
   c) check whether there are B/F nodes and/or links that can be deactivated in order to save energy;
   d) if so, update the current network configuration accordingly.

Sec. IV-A explains how we obtain the initial solution, i.e., Item 1 above. Items (a)–(b) and (c)–(d) correspond to the `fixProblems` and `saveEnergy` procedures respectively, which are described in Sec. IV-B and Sec. IV-C.

It is worth pointing out that the `fixProblems` and `saveEnergy` procedures are designed to take no action if no action is warranted, and therefore there is no harm in cascading them. As an example, `fixProblems` will never take any action the first time it is executed after an initial solution is generated, as that solution is guaranteed to be feasible. Similarly, `saveEnergy` is unlikely to find elements to deactivate if `fixProblems` just had to activate some.

### A. Initial solution

The initial solution used to initialize OptiLoop has to be feasible, but does not have to be optimal. It can come from one of the heuristics we reviewed in Sec. II, or it can be obtained by solving a version of our problem where:

1) all B/F nodes and links are active, i.e., $y_c = 1, \forall c \in \mathcal{C}$ and $x_{i,j} = 1, \forall (i,j) \in \mathcal{L}$;
2) there is an instance of all VNFs deployed at each B/F node, i.e., $\delta(c,v) = 1, \forall c \in \mathcal{C}, v \in \mathcal{V}$.

The resulting solution will be highly suboptimal, as we are likely to needlessly activate B/F nodes and/or links and to place useless VNF instances, all of which increase the power consumption. On the plus side, the problem is LP, as all binary variables are fixed; furthermore, the following property holds.

*Property 1:* If a problem instance is feasible, then there is at least one feasible solution where the $x, y$ and $\delta$ variables are all set to 1.

Please refer to [16] for the proof.

In other words, setting all binary variables to one is an easy way to obtain a feasible solution to our problem to start with. This solution can be vastly improved, as discussed next.

### B. The `fixProblems` procedure

The high-level goal of the `fixProblems` procedure is to check whether the current network configuration can cope with the current (and projected) traffic demand. If this is not the case, then we take one or more of the following actions: (i) activating additional B/F nodes; (ii) activating additional links; (iii) deploying additional VNF instances.

**Algorithm 1** The `fixProblems` procedure.
**Require:** $\mathcal{S}^{\text{curr}}$
1: $\mathcal{P} \leftarrow$ **new** problem()
2: $\mathcal{P}.\text{fix}(x_{i,j} \leftarrow x_{i,j}^{\text{curr}}, \quad \forall (i,j) \in \mathcal{L})$
3: $\mathcal{P}.\text{fix}(y_c \leftarrow y_c^{\text{curr}}, \quad \forall c \in \mathcal{C})$
4: $\mathcal{P}.\text{fix}(\delta(c,v) \leftarrow \delta^{\text{curr}}(c,v), \quad \forall c \in \mathcal{C}, v \in \mathcal{V})$
5: **solve**$(\mathcal{P})$
6: **if** $\mathcal{P}.\text{is\_feasible}$ **then**
7:     **return**
8: **if** $(4) \in \mathcal{P}.\text{IIS}$ **then**
9:     $\mathcal{P}.\text{relax}(x_{i,j}: x_{i,j}^{\text{curr}} = 0)$
10:     $\mathcal{P}.\text{relax}(y_c: y_c^{\text{curr}} = 0)$
11:     $\tilde{x}, \tilde{y} \leftarrow$ **solve**$(\mathcal{P})$
12:     $(i^\star, j^\star) \leftarrow$ **choose** from $\mathcal{L}$ with prob. $\tilde{x}_{i,j}$
13:     $\mathcal{P}.\text{fix}(x_{i^\star, j^\star} \leftarrow 1)$
14:     $\mathcal{P}.\text{fix}(y_i \leftarrow 1; y_j \leftarrow 1)$
15:     **goto** Line 5
16: **if** $(7) \in \mathcal{P}.\text{IIS}$ **then**
17:     $\mathcal{P}.\text{relax}(y(c): y^{\text{curr}}(c) = 0)$
18:     $\mathcal{P}.\text{relax}(\delta(c,v): \delta^{\text{curr}}(c,v) = 0)$
19:     $\tilde{\delta} \leftarrow$ **solve**$(\mathcal{P})$
20:     $c^\star, v^\star \leftarrow$ **choose** from $\mathcal{C} \times \mathcal{V}$ with prob. $\tilde{\delta}(c,v)$
21:     $\mathcal{P}.\text{fix}(y(c^\star) \leftarrow 1)$
22:     $\mathcal{P}.\text{fix}(\delta(c^\star, v^\star) \leftarrow 1)$
23:     **goto** Line 5

Specifically, as detailed in Alg. 1, we take as an input the current solution $\mathcal{S}^{\text{curr}}$. We then proceed, in Line 1–Line 4, to create a new instance $\mathcal{P}$ of the problem, where all binary variables are fixed to their values in $\mathcal{S}^{\text{curr}}$. In Line 5, we solve such a problem: if it is feasible, then no action is required and the algorithm exits (Line 7). Otherwise, we look at why the problem is unfeasible, by inspecting its *irreducible inconsistent subsystem* (IIS), i.e., the subset of constraints such that removing any of them would make the problem feasible. This set allows us to discriminate between the different reasons that can make the network unable to operate properly (hence, the problem unfeasible).

If constraint (4) (mandating that no link is used for more than its capacity) is in the IIS, then we need to activate some more links and/or B/F nodes. To decide which ones, we relax all $x$- and $y$-variables related to B/F nodes and links that were inactive in $\mathcal{S}^{\text{curr}}$ (Line 9–Line 10) and solve the new problem (Line 11). We then choose one link to activate, with a probability proportional to its relaxed $\tilde{x}_{i,j}$ value, and fix to 1 the corresponding $x$-value and the $y$-values of its endpoints (Line 12–Line 14). We then go back to Line 5 and test the new solution (Line 15). If it is still infeasible, we will activate further network elements until feasibility is achieved.

We proceed in a similar way if constraint (7) is in the IIS, i.e., if we have a computational capability issue. We relax variables $y$ and $\delta$, allowing for more B/F nodes to be activated and VNFs to be deployed if needed, and solve the new problem obtaining the relaxed values $\tilde{\delta}$ (Line 17–Line 19). We then have to decide which VNF to place and where. We do so by selecting a B/F node $c^\star$ and a VNF $v^\star$ at random, with a probability proportional to the relaxed values $\tilde{\delta}(c,v)$, and fix the corresponding $y$ and $\delta$-variable to 1 (Line 20–Line 22). Finally, we go back to testing the new solution (Line 23).

Note that all problems we solve in Alg. 1 are LP: in Line 5, Line 11 and Line 19 all binary variables are either fixed or relaxed. Such problems can be therefore solved in polynomial time (*embedded* [17] optimization on low-power hardware is now commonplace in several application domains).

### C. The saveEnergy procedure

We can think of the saveEnergy procedure as the dual of fixProblems. Our aim is to identify B/F nodes and/or links that can be deactivated, as well as VNF instances that can be removed from the B/F nodes they run into. The objective is to reduce our power consumption without impairing our ability to serve the traffic, i.e., without making the problem infeasible. As in the fixProblems procedure, we solve a sequence of LP problems with fixed or relaxed variables, obtaining guidance on the decisions we should make and their effects.

In Alg. 2, we take the current solution $\mathcal{S}^{\text{curr}}$ as an input. We then create an instance $\mathcal{P}$ of the problem where the binary variables that in the current solution have value 0 are fixed to 0 (Line 2–Line 4), and those that have currently value 1 are relaxed (Line 5–Line 7). This is because we are not looking for new nodes/links to activate, but for elements to deactivate. We do so by solving the problem instance $\mathcal{P}$ (Line 8); note that all binary variables therein are fixed or relaxed, so the problem is LP.

In Line 9–Line 11 we identify the link, B/F node, and pair of B/F node and VNF that are active in the current solution and have the lowest value of the associated relaxed variable (respectively $\tilde{x}_{i,j}$, $\tilde{y}(c)$, and $\tilde{\delta}(c,v)$). Intuitively, these are the elements that most likely can be deactivated without impairing network functionality. We check this by creating a copy of problem instance $\mathcal{P}$ and fixing to 0 the binary variable associated to the element with the lowest value of the relaxed variables (Line 12–Line 20). If that element is a B/F node, we also need to deactivate the links using it and the VNF instances it hosts (Line 17–Line 18).

The difference between $\mathcal{P}$ and $\mathcal{P}_2$ is that exactly one element that was active in $\mathcal{P}$ is deactivated in $\mathcal{P}_2$, hence $\mathcal{P}_2$ is also LP. In Line 21, we solve $\mathcal{P}_2$ and check if it is feasible. If that is the case, then we use $\mathcal{P}_2$ as our new solution, and try to further enhance it (Line 23–Line 24). Otherwise, the algorithm returns $\mathcal{P}$, the last feasible solution we tried.

In summary, Alg. 2 deactivates zero or more elements, i.e., B/F nodes, links, or VNF instances. The element to deactivate is chosen based on the value taken by the corresponding relaxed variable, and after each change we check that the resulting configuration can serve its load, i.e., the problem instance is feasible.

### V. TESTBEDS, SCENARIO AND BENCHMARKS

We validate and evaluate OptiLoop through two testbeds. We study the interaction between OptiLoop, the SDN controller, and the NFVO in a small-scale testbed with real hardware, described in Sec. V-A. For our performance evaluation we instead use a larger, emulated testbed based on the real-world topology of a mobile operator, as detailed in Sec. V-B.

### A. OpenDaylight/OpenStack testbed

The architecture of our testbed is summarized in Fig. 4(left). OpenDaylight (Beryllium version) and OpenStack (Mitaka version) are used to control a network made of three Lagopus

---

**Algorithm 2** The saveEnergy procedure.

**Require:** $\mathcal{S}^{\text{curr}}$
1: $\mathcal{P} \leftarrow$ **new** problem()
2: $\mathcal{P}.\text{fix}(x_{i,j} \leftarrow 0, \quad \forall (i,j) \in \mathcal{L}: x_{i,j}^{\text{curr}} = 0)$
3: $\mathcal{P}.\text{fix}(y_c \leftarrow 0, \quad \forall c \in \mathcal{C}: y_c^{\text{curr}} = 0)$
4: $\mathcal{P}.\text{fix}(\delta(c,v) \leftarrow 0, \quad \forall c \in \mathcal{C}, v \in \mathcal{V}: \delta(c,v) = 0)$
5: $\mathcal{P}.\text{relax}(x_{i,j}, \quad \forall (i,j) \in \mathcal{L}: x_{i,j}^{\text{curr}} = 1)$
6: $\mathcal{P}.\text{relax}(y_c, \quad \forall c \in \mathcal{C}: y_c^{\text{curr}} = 1)$
7: $\mathcal{P}.\text{relax}(\delta(c,v), \quad \forall c \in \mathcal{C}, v \in \mathcal{V}: \delta(c,v) = 1)$
8: **solve**($\mathcal{P}$)
9: $(x^\star, y^\star) \leftarrow \arg\min_{(x,y) \in \mathcal{L}: x_{x,y}^{\text{curr}} = 1} \tilde{x}_{i,j}$
10: $c^\star \leftarrow \arg\min_{c \in \mathcal{C}: y^{\text{curr}}(c) = 1} \tilde{y}(c)$
11: $d^\star, v^\star \leftarrow \arg\min_{c,v \in \mathcal{C} \times \mathcal{V}: \delta^{\text{curr}}(c,v) = 1} \tilde{\delta}(c,v)$
12: $\mathcal{P}_2 \leftarrow$ **copy**($\mathcal{P}$)
13: **if** $\tilde{x}_{i^\star,j^\star} < \tilde{y}(c^\star) \wedge \tilde{x}_{i^\star,j^\star} < \tilde{\delta}(d^\star,v^\star)$ **then**
14: $\quad \mathcal{P}_2.\text{fix}(x_{i^\star,j^\star} \leftarrow 0)$
15: **if** $\tilde{y}(c^\star) < \tilde{x}_{i^\star,j^\star} \wedge \tilde{y}(c^\star) < \tilde{\delta}(d^\star,v^\star)$ **then**
16: $\quad \mathcal{P}_2.\text{fix}(y(c^\star) \leftarrow 0)$
17: $\quad \mathcal{P}_2.\text{fix}(x_{i,j} \leftarrow 0, \quad \forall (i,j) \in \mathcal{L}: i = c^\star \vee j = c^\star)$
18: $\quad \mathcal{P}_2.\text{fix}(\delta(c,v) \leftarrow 0, \quad \forall c \in \mathcal{C}, v \in \mathcal{V}: c = c^\star)$
19: **if** $\tilde{\delta}(d^\star,v^\star) < \tilde{x}_{i^\star,j^\star} \wedge \tilde{\delta}(d^\star,v^\star) < \tilde{y}(c^\star)$ **then**
20: $\quad \mathcal{P}_2.\text{fix}(\delta(d^\star,v^\star) \leftarrow 0)$
21: **solve**($\mathcal{P}_2$)
22: **if** $\mathcal{P}_2.\text{is\_feasible}$ **then**
23: $\quad \mathcal{P} \leftarrow \mathcal{P}_2$
24: $\quad$ **goto** Line 1
25: **else**
26: $\quad$ **return** $\mathcal{P}$

---

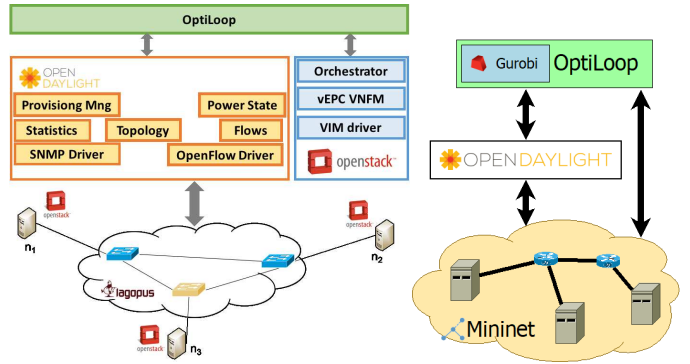

Fig. 4. Architecture of our testbeds: real hardware (left) and emulated (right). In both testbeds, OptiLoop is implemented as standalone application. In the real testbed, it controls a network composed of three software switches running Lagopus and three B/F nodes $n_1 \ldots n_3$. In the emulated testbed, it controls a Mininet-emulated topology reproducing the network of a real-world operator.

software switches (with DPDK support enabled for faster switching) and three physical servers. The OpenDaylight SDN controller configures the data plane, by activating/deactivating links and switches via SNMP protocol and configuring the forwarding rules via OpenFlow 1.3 protocol. A custom-built NFVO – integrated with the VNFM (VNF manager) and VIM (Virtual Interface Manager) OpenStack modules – manages the VMs that run the VNFs. We adopt the OpenAirInterface [14] vEPC implementation, including the four VNFs in Fig. 1.

OptiLoop is implemented as a standalone application, written in Java and including two main components, devoted to monitoring and decision-making. OptiLoop interacts with both OpenDaylight and the NFVO through their REST APIs, gathering up-to-date information on the status of switches, links, physical servers and VNFs. When a decision is made, it communicates it to OpenDaylight (if the decision concerns link activation/deactivation) or the NFVO (if the decision concerns VNF deployment or server activation/deactivation). The decision-making component essentially implements Alg. 1 and Alg. 2, using the Gurobi solver for optimization. Since Gurobi features Java bindings, using it within the OptiLoop application is as simple as importing a library.

### B. Emulated testbed

Our performance evaluation is carried out through an emulated testbed based on Mininet, the *de facto* standard solution to study SDN-based networks. Its architecture is summarized in Fig. 4(right): similarly to the previous case, OptiLoop interacts with the OpenDaylight controller for network management, and directly with Mininet via its Python API to turn servers and switches on and off. Notice that the actual VNFs are *not* implemented in Mininet; the traffic they serve is emulated via `iperf` and the energy consumption is estimated via realistic models, as detailed in Sec. V-B2.

The switches and servers emulated by Mininet reproduce the real-world topology of a major mobile operator, as detailed in Sec. V-B1. Links and servers are implemented through the `TCLink` and `CPULimitedHost` Mininet classes, which allow us to assign them bandwidth, delay and computational capability matching those of their real-world counterparts. All `iperf`-generated traffic is based on the real-world traffic figures we have access to.

*1) Network topology and traffic:* Our reference topology, displayed in Fig. 5, represents the real-world topology of a major mobile network operator. It includes 42 endpoints and 51 B/F nodes, with each endpoint connected to exactly two B/F nodes. A total of 1,497 antennas are connected to the endpoints. Accounting for the expected future traffic growth [18], we have an aggregate traffic varying between $74\,\mathrm{Mbit/s}$ and $473\,\mathrm{Mbit/s}$ per endpoint, with a 82:18 downlink/uplink proportion. The dataset we use only represents a snapshot of the network conditions, i.e., traffic demand does not change over time.

Based on the real-world vEPC implementation [14] we consider a total of four VNFs, namely eNB, MME, HSS, and a gateway implementing both the P-GW and S-GW functions.

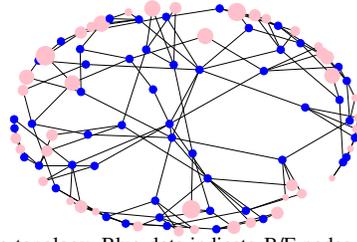

Fig. 5. Reference topology. Blue dots indicate B/F nodes, pink ones indicate endpoints. The size of pink dots is proportional to the traffic they generate.

Notice that in [14] no VNF is split into user- and control-plane sub-entities. We set our $\chi$-values, expressing how traffic gets transformed as it travels between VNFs, leveraging the analysis in [10]; in particular, the fraction of control traffic going to the MME is given by $\chi(\mathrm{eNB}, \mathrm{P/S\text{-}GW}, \mathrm{MME}) = 0.32$.

Still based on [10], we set the link bandwidth $C_{i,j}$ to $10\,\mathrm{Gbit/s}$ for endpoint-to-node links and $100\,\mathrm{Gbit/s}$ for node-to-node ones. Based on [10] and [19], we assume that each B/F node can process $100\,\mathrm{Gbit}$ of traffic every second. Since our scenario is constrained by B/F node and link capacity, we ignore network and processing delays.

*2) Power consumption figures: Idle power* is the power consumed by any active B/N node, no matter how much it is used; in our model, it is indicated as $f_{\mathrm{idle}}$. Based on [20], we set that power to $65\,\mathrm{W}$. Upon deploying a VNF at a B/F node, we might expect to incur an *allocation energy* cost due to, for example, the overhead of the VM or container. However, the measurements [20] show that such overhead is negligible, i.e., we can set $f_0 = 0$.

Assuming a software switch running Lagopus, *switching* $40\,\mathrm{Gbit/s}$ fully utilizes an Intel Xeon E5-2680 processor [19], which has a $130\,\mathrm{W}$ thermal design power. This gives us a figure of $f_{\mathrm{sw}}(c) = 3.25\,\mathrm{nJ/bit} \cdot \sum_{e,v_1,v_2} t_c(e, v_1, v_2)$. $f_{\mathrm{link}}$, instead, can be neglected [20]. Using the computational capability figures from [10] and assuming a high-end, 32-core processor, we set $f_{\mathrm{proc}}(c) = 48\,\mathrm{nJ/bit} \cdot \sum_{e,v_1,v_2} p_c(e, v_1, v_2)$.

It is worth stressing that all these energy consumption figures are subject to rapid change as technology evolves, and that they represent only an input to our problem.

*3) Benchmark solutions:* We compare OptiLoop with two alternatives: what *is* done in current real-world systems, i.e., keeping all network elements active regardless of traffic conditions, and state-of-the-art approaches from the literature.

Among the VNF placement schemes reviewed in Sec. II, we select the popular approach of *consolidation* (used, e.g., in [7]) as our benchmark. The consolidation procedure consists of three-stage decision process. For every flow, it first looks for an already-deployed VNF to serve the flow; if none can be found, it deploys a new instance of the VNF at an already active B/F node. If no suitable node is found, it activates a new one. Also, the procedure activates any additional B/F nodes needed to ensure connectivity between endpoints and the serving B/F nodes.

## VI. RESULTS

We start this section by summarizing our validation results from the OpenDaylight/OpenStack testbed, in Sec. VI-A.

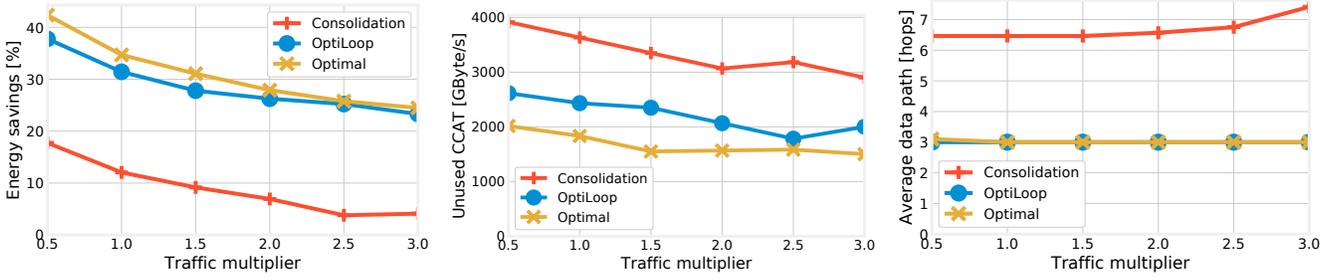

Fig. 6. Energy savings obtained as a function of traffic (left); spare computational capabilities of the active topology (CCAT) (center); number of hops traveled by requests (right).

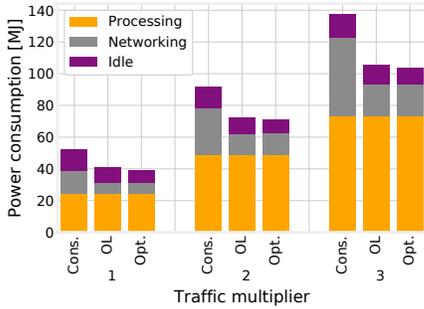

Fig. 7. Breakdown of energy consumption for the consolidation-based ("Cons."), OptiLoop ("OL"), and optimal ("Opt.") strategies.

We then present, in Sec. VI-B, a performance evaluation of OptiLoop carried out by emulating a real-world topology in Mininet, as described in Sec. V-B.

### A. Validation results

Our validation has a twofold purpose. The first is to ensure that the interaction between OptiLoop, OpenStack and OpenDaylight – both information retrieval and decision enacting – works as intended. The second is measuring the *delay* resulting from OptiLoop's decisions. As we can see from Tab. II, the delays of powering on/off servers and switches and activating/deactivating links are quite low. Conversely, instantiating a new VNF can require over a minute, mainly due to the time it takes to start a new VM.

It is however interesting to note that the time to instantiate the full service, i.e., all four VNFs, is only slightly higher than the time to instantiate one VNF, as the NFVO can handle multiple VNFs in parallel. This suggests that our system can scale to complex services requiring many VNFs. Also, the delays in Tab. II suggest that containers could represent an attractive alternative to VMs, due to their shorter setup times.

### B. Emulation-based performance evaluation

The first answer we seek from the performance evaluation carried out through the emulated testbed concerns the magnitude of possible energy savings. In Fig. 6(left), we vary the traffic demand between 0.5 and 3 times the real-world amount, and study how much energy we can save compared to what is done today, i.e., leaving all B/F nodes and links active. We can observe that OptiLoop yields dramatic savings, consistently very close to the optimum, while consolidation does not perform as well. An intuitive reason is that OptiLoop accounts for all the three main contributions to energy consumption (processing, idle power, and networking), while the consolidation-based approach focuses on keeping the number of active B/F nodes low.

TABLE II
DRIVING REAL HARDWARE: INCURRED DELAYS [S]

| Component | Maximum | Minimum | Average |
|---|---|---|---|
| Server power on | 0.021 | 0.010 | 0.016 |
| Switch power on | 0.039 | 0.019 | 0.026 |
| Link activation | 4.246 | 3.556 | 3.786 |
| VNF instantiation | 60.442 | 40.222 | 54.559 |
| Service instantiation | 82.066 | 62.356 | 64.710 |

Fig. 6(center) shows the spare computational capability of the active topology (CCAT); intuitively, this is a measure of how much power is being wasted, i.e., how inefficient the network management strategy is. The consolidation algorithm has the highest spare CCAT, because of the higher number of B/F nodes that have to be activated in order to guarantee connectivity. The spare CCAT yielded by OptiLoop is much lower, and very close to the optimum. It is interesting to remark that even the optimum leaves substantial spare CCAT. This is due to the fact that some B/F nodes have to be active in order to keep the topology connected, even if they do not have to host any VNF. Fig. 6(right) depicts how many hops data travels across the network. OptiLoop again matches the optimum, while the consolidation strategy results in substantially longer paths, due to the fact that VNF placement decisions are made without accounting for connectivity.

Fig. 7 breaks the total energy consumption into its main components, namely, processing, networking, and idle power. Note that these components have comparable magnitude, i.e., none of them dominates the overall consumption. It follows that network management strategies have to account for them all. We can also see that the processing component never changes across strategies, since the amount of traffic to process is always the same. The difference between the strategies lies mostly in the networking component (longer paths in Fig. 6(right) correspond to higher consumption) and, to a lesser extent, in the idle energy. In other words, it is important to place VNFs close to the traffic they have to serve, while at the same time activating as few B/F nodes as possible.

Fig. 8(left) and Fig. 8(center) show that placing VNFs close to the traffic they serve also means placing *many* of them. This goes against the traditional concept of activating only the strictly required number of elements, and it is a direct consequence of the features of modern, software-based networks. As discussed in Sec. V-B2, there is little or no penalty for placing an underutilized VNF instance on an already active B/F node, while there is a significant energy cost for transferring even modest amounts of data between B/F nodes. Indeed, we can say that OptiLoop outperforms state-of-the-art alternatives *because* it properly accounts for the unique features of 5G, thus being more aggressive in deploying VNFs.

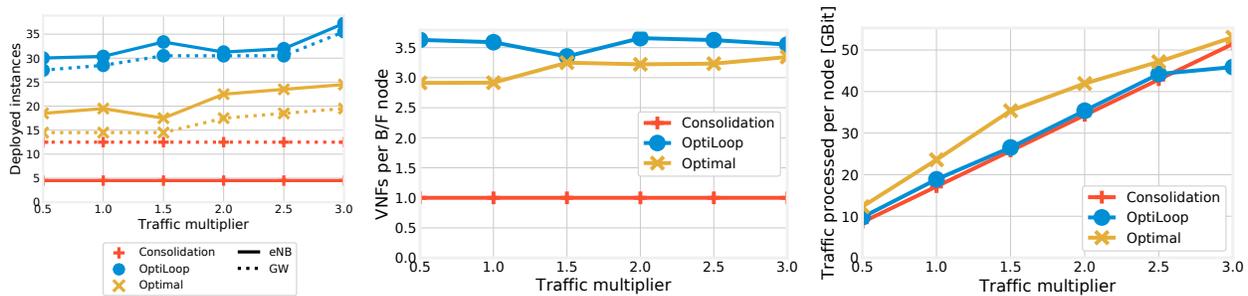

Fig. 8. Number of deployed instances for the eNB and P/S-GW VNFs (left); average number of VNFs deployed in each B/F node (center); average traffic processed at each B/F node (right).

Comparing Fig. 8(left) to Fig. 8(center), we can see that OptiLoop deploys more VNFs than the optimum, but the number of VNFs per B/F node is similar. This is because OptiLoop activates slightly more B/F nodes than the optimum, as confirmed by Fig. 8(right) showing that the average amount of traffic processed per B/F node is slightly lower in OptiLoop.

## VII. CONCLUSIONS

We considered 5G networks where nodes may have both networking and computation capabilities. We addressed the problem of the energy-efficient management of such networks, requiring to decide (i) which B/F nodes and links to activate, (ii) how many instances to deploy for each VNF, and how to place them across the active B/F nodes, and (iii) how to route traffic between them. We proposed an optimization model accounting for the unique features of 5G traffic, and used it to formulate OptiLoop: a real-time, efficient network management strategy based on repeatedly solving relaxed optimization problems.

We validated OptiLoop through a testbed including the OpenDaylight SDN controller and the OpenStack virtualization infrastructure. We then evaluated its performance through network emulation, with both the network topology and the traffic demand coming from a real-world dataset provided by a major network operator. We found that OptiLoop outperforms state-of-the-art alternatives and performs very close to the optimum, mainly thanks to its ability to account for all the main sources of energy consumption that characterize 5G systems.

## ACKNOWLEDGEMENT

This work was supported by the EU project "5G- Crosshaul: The 5G Integrated fronthaul/backhaul" (grant no. 671598) within the H2020 programme.